\documentclass[english,aps,prl,twocolumn,showpacs,amsmath,amssymb]{revtex4}
\usepackage{graphicx}
\usepackage{color}
\usepackage{amsmath}

\begin{document}
\title{Singular diffusion and criticality in a confined sandpile}

\author{R.~S.~Pires}
\author{A.~A.~Moreira}
\author{H.~A.~Carmona}
\author{J.~S.~Andrade~Jr.}
\affiliation{Departamento de F\'{\i}sica, Universidade Federal do
  Cear\'a, 60451-970 Fortaleza, Cear\'a, Brazil}

\date{\today}

\begin{abstract}
  We investigate the behavior of a two-state sandpile model subjected
  to a confining potential in one and two dimensions. From the
  microdynamical description of this simple model with its intrinsic
  exclusion mechanism, it is possible to derive a continuum nonlinear
  diffusion equation that displays singularities in both the diffusion
  and drift terms. The stationary-state solutions of this equation,
  which maximizes the Fermi-Dirac entropy, are in perfect agreement
  with the spatial profiles of time-averaged occupancy obtained from
  model numerical simulations in one as well as in two dimensions.
  Surprisingly, our results also show that, regardless of dimensionality,
  the presence of a confining potential can lead to the emergence of
  typical attributes of critical behavior in the two-state sandpile
  model, namely, a power-law tail in the distribution of avalanche sizes.
\end{abstract}

\maketitle

Physical processes involving anomalous diffusion are typically
associated with systems in which the mean square displacement of their
elementary units follows a nonlinear power-law relationship with time,
$\sigma^{2}\propto t^{\alpha}$, with an exponent $\alpha
\neq 1$, in contrast with linear standard diffusion ($\alpha=1$). Instead of
being a rare phenomenon, as suggested by its own denomination, anomalous
diffusion, however, appears rather ubiquitously in Nature, playing an important
role in a variety of scientific and technological applications, such as fluid
flow through disordered porous media~\cite{Lukyanov2012}, surface
growth~\cite{Spohn1993}, diffusion in fractal-like
substrates~\cite{Stephenson1995,Andrade1997,Buldyrev2001,Costa2003,Havlin2002},
turbulent diffusion in the atmosphere~\cite{Richardson1926,Hentschel1984},
spatial spreading of cells~\cite{Simpson2011} and biological
populations~\cite{Colombo2012}, cellular transport~\cite{Caspi2002}, and
cytoplasmic crowding in cells~\cite{Weiss2004}. Anomalous diffusion can also
manifest its non-Gaussian behavior in terms of nonlinear Fokker-Plank
equations~\cite{Lenzi2001,Malacarne2001,Malacarne2002,DaSilva2004,Lenzi2005},
which is the case, for example, of the dynamics of interacting vortices in
disordered superconductors~\cite{Zapperi2001,Moreira2002,Miguel2003,Andrade2010},
diffusion in dusty plasma~\cite{Liu2008,Barrozo2009}, and pedestrian
motion~\cite{Barrozo2009}.

The extreme case of nonlinear behavior in diffusive systems certainly
corresponds to {\it singular diffusion}. For instance, in some
physical conditions, the diffusion of adsorbates on a surface can be
strongly nonlinear~\cite{Ehrlich1980,Gomer1990,Myshlyavtsev1995}, with
a surface diffusion coefficient that depends on the local coverage
$\theta$ as, $D\propto{\left|\theta-\theta_{c}\right|^{-\alpha}}$.
The study of surface-diffusion mechanisms is crucial for the
understanding of technologically important processes related with
physical adsorption~\cite{Vidali1991} and catalytic surface
reactions~\cite{Manandhar2003,Hofmann2005,Schmidtbauer2012}. In
particular, a singularity in the coverage dependence of the diffusion
coefficient is frequently associated to continuous phase
transitions~\cite{Myshlyavtsev1995}.

A direct connection between singular diffusion and self-organized
criticality~\cite{Bak1987} has been disclosed by Carlson {\it et
  al.}~\cite{Carlson1990,Carlson1993} in terms of a two-state
one-dimensional sandpile model with a driving mechanism, where grains
are added at one end of the pile and fall off at the other end.
Besides exhibiting a self-organized state, the continuum limit of this
simple model leads to a nonlinear diffusion equation, where the
diffusion coefficient not only depends on the local density, but also
displays a singularity at a ``critical'' density
value~\cite{Carlson1990,Carlson1993,Carlson1995,Kadanoff1992,Barbu2010}.
Indeed, the critical aspects of this model remain to be elucidated,
specially due to the fact that the most prominent sign of criticality,
namely, long-range power-law spatial correlations are not present in
the original setup of the simulated dynamical system. Here we show
that the addition of a confining potential to the two-state sandpile
model solves this problem, namely, power-law tails are observed in the
distribution of avalanche sizes in both one- and two-dimensional
versions of the theoretical model. Moreover, our results reveal that
the continuum description of the model contains singular
nonlinearities in both the diffusion and drift terms of the resulting
partial differential equation for the transport process.

The microscopic model investigated in this study consists of an
one-dimensional lattice of size $N_{s}$ on which $N$ particles
($N_{s}\gg N$) are randomly placed in such a way that the height
$h(i)$ of each site $i=1,\,\dots,\, N_{s}$ is either $1$ or $0$. At
each step, one grain is chosen randomly to move to the left or to the
right with equal probability. If the nearest neighbor in the chosen
direction is occupied, the grain jumps instantly to the next-nearest
neighbor in the same direction. If this site is also occupied, the
particle keeps jumping until it finally reaches an empty site
$j$~\cite{Carlson1990}. This type of exchange driving mechanism for
closed systems has been previously introduced in the context of
fluctuations and local equilibrium in self-organizing
systems~\cite{Carlson1993,Montakhab1998}. Here, an external confining
potential is applied to the system by introducing a non-uniform
transition probability from site $i$ to $j$. Precisely, each site is
mapped into the continuous interval $[-L/2,L/2]$, and the position
$x_{i}=i(L/N_{s})-L/2$ is associated with a potential energy
$\phi(x_{i})$. For a given transition, we compute
$\Delta\phi_{ij}=\phi(x_{j})-\phi(x_{i})$ and use the following
Metropolis rules:
\begin{align*}
   \left.\begin{array}{c}
     h(i)\rightarrow h(i)-1\\
     h(j)\rightarrow h(j)+1
   \end{array}\right\}  & 
   \text{ if }\Delta\phi_{ij}<0\text{ or } r< w=\exp\left(-\beta\Delta\phi_{ij}\right)\\
   \left.\begin{array}{c}
     h(i)\rightarrow h(i)\\
     h(j)\rightarrow h(j)
   \end{array}\right\}  & \text{ if }\Delta\phi_{ij}>0\text{ and }r>w, 
\end{align*}
where $r$ is a uniform random number in the interval $[0,1]$,
$\beta\equiv 1/k_{B}T$, $T$ is the temperature of the thermal
reservoir in contact with the system, $k_{B}$ is the Boltzmann
constant, and we count one unit of time for every $N$ grains moved.
The effect of decreasing the temperature is equivalent to increasing
the strength of the external potential.
\begin{figure}[t]
	\includegraphics[width=\columnwidth]{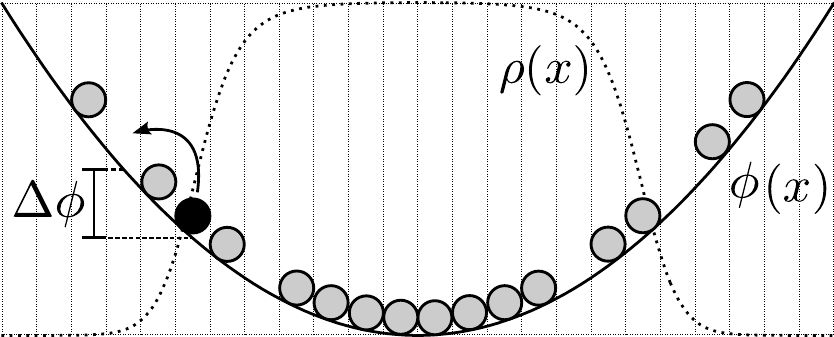}
	\caption{Illustration of the model. A grain moves if
          $\Delta\phi_{ij}<0$ or $r<e^{-\beta\Delta\phi_{ij}}$,
          $r\in[0,1]$. As a result, the average density $\rho$ has a
          maximum in the region where $\phi$ is minimum.}
      \label{fig:model}
\end{figure}

A continuum limit for this microscopic model can be obtained
rigorously. If we let $\rho_{i}=\rho(x_{i},\,t)$ be the probability
that site $i$ located at $x_{i}$ is occupied at time $t$ and
$\phi_{i}=\phi(x_{i})$, a master equation can then be written as,
\begin{widetext}
	\begin{multline}
          \frac{\partial\rho_{i}}{\partial
            t}=-\frac{\rho_{i}}{\tau}\left\{ \frac{1}{2}
            \sum_{j=1}^{\infty}(1-\rho_{i+j})\min[1,e^{-\beta(\phi_{i+j}-\phi_{i})}]
            \prod_{k=1}^{j-1}\rho_{i+k}+
            \frac{1}{2}\sum_{j=1}^{\infty}(1-\rho_{i-j})\min[1,e^{-\beta(\phi_{i-j}-\phi_{i})}]
            \prod_{k=1}^{j-1}\rho_{i-k}\right\} \\
          +\frac{(1-\rho_{i})}{\tau}\left\{
            \frac{1}{2}\sum_{j=1}^{\infty}\min[1,e^{-\beta(\phi_{i}-\phi_{i+j})}]\prod_{k=1}^{j}
            \rho_{i+k}+\frac{1}{2}\sum_{j=1}^{\infty}\min[1,e^{-\beta(\phi_{i}-\phi_{i-j})}]
            \prod_{k=1}^{j}\rho_{i-k}\right\},
          \label{masterEq}
	\end{multline}
\end{widetext}
where $\tau$ is the average time between transitions. The first term
on the right side is the transition rate corresponding to site $i$
being occupied at time $t$ and loosing the grain, while the second
term accounts for the transition rate for an empty site $i$ to gain a
grain. Considering that $\tau\approx\delta^{2}/2D$, where $\delta$ is
the lattice spacing and $D>0$ is a constant with dimensions of
diffusion coefficient ($cm^2/s$), and keeping terms to order
$\mathcal{O}(\delta^{2})$, it can be shown that, as $\delta$ goes to
zero, the following nonlinear diffusion equation holds:
\begin{equation}
  \frac{\partial\rho}{\partial t}=D\frac{\partial}{\partial x}\left[\frac{(1+\rho)}{(1-\rho)^{3}}
  \frac{\partial\rho}{\partial x}+\frac{(1+\rho)}{(1-\rho)^{2}}\beta\frac{d\phi}{dx}\rho\right].
  \label{diffusionEq}
\end{equation}
Details of this derivation can be found in the Supplementary Material.
Equation~(\ref{diffusionEq}) can be related to a nonlinear
Fokker-Planck equation (FPE) of the form,
\begin{equation}
  \frac{\partial\rho}{\partial t}=
  \frac{\partial}{\partial x} \left[\Omega(\rho)\frac{\partial\rho}{\partial x}\right]
 -\frac{\partial}{\partial x} \left[A(x)\Psi(\rho)\right],
  \label{FPE_Eq}
\end{equation}
with $\Omega(\rho)=D(1+\rho)/(1-\rho)^{3}$, $A(x)=-d\phi(x)/dx$, and
$\Psi(\rho)=D\beta\rho(1+\rho)/(1-\rho)^{2}$. Considering the
FPE~(\ref{FPE_Eq}), with $dF/dt\le0,$where $F=U-\gamma S,$ $U=\int
dx\rho(x,t)\phi(x)$, and the entropy taken in a general form as
$S[\rho]=\int dx\,g[\rho(x)]$, with $g(0)=g(1)=0$ and $d^{2}g/d\rho^{2}\le0$, 
we obtain~\cite{Schwammle2007,Andrade2010},
\begin{equation}
  -\gamma\frac{d^{2}g(\rho)}{d\rho^{2}}=\frac{\Omega(\rho)}
  {\Psi(\rho)}=\frac{1}{\beta\rho(1-\rho)}, \label{d2gdp2}
\end{equation}
where $\gamma$ is a positive Lagrange multiplier. This equation has a
solution in the form,
\begin{equation}
  g(\rho)=\frac{-\rho\ln\rho-(1-\rho)\ln(1-\rho)}{\beta\gamma},
  \label{entropy}
\end{equation}
for which the entropy $S[\rho]=\int dx\, g[\rho(x)]$ reduces to the
entropy of a Fermi gas, with $\gamma=T$. The functional $\Omega(\rho)$
physically corresponds to a diffusion coefficient which depends on
$\rho(x,\,t)$. Clearly it diverges for $\rho=1$ and the diffusion
coefficient has the same form as for the case without the external
potential \cite{Carlson1990}. The functional $\Psi(\rho)$ is related
to a drift due to the external potential, and also diverges for
$\rho=1$.

\begin{figure}[t]
\includegraphics[width=\columnwidth]{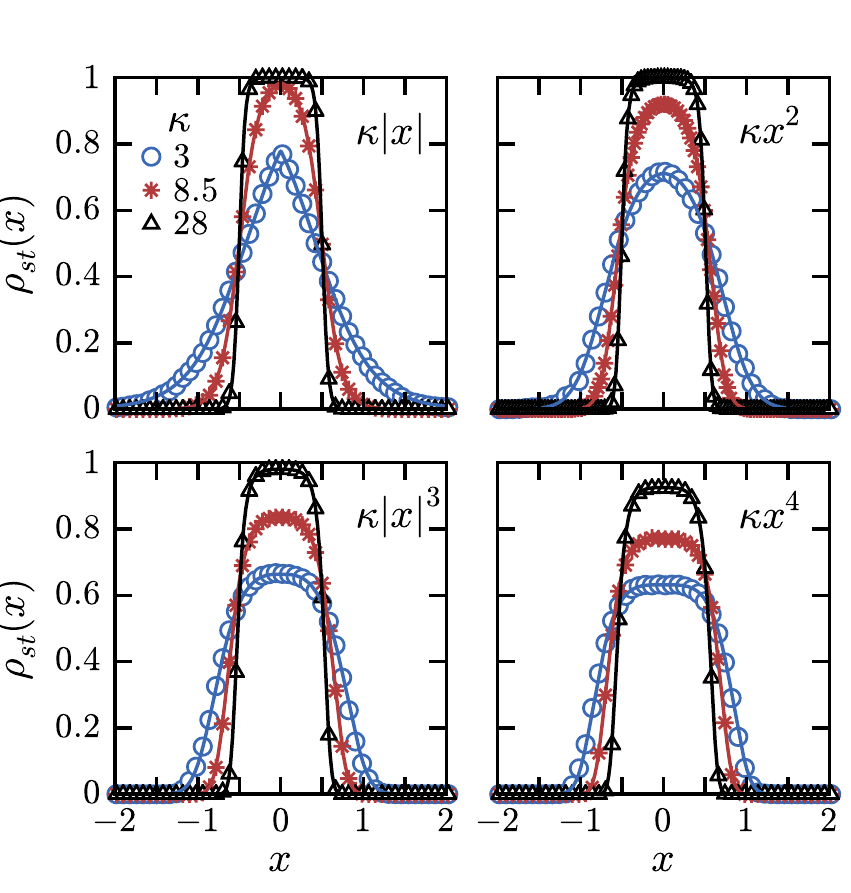}
\caption{Comparison between numerical and analytical stationary-state
  solutions for the occupancy density of the confined two-state
  sandpile model in 1D.  Numerical results are for a system with
  $N=4000$ grains, $\beta=1$, and potentials given by
  $\phi(x)=\kappa{|x|}^{n}$, with $n=1,\,2,\,3$ and $4$. The $\kappa$
  values are $3.0$ (blue circles), $8.5$ (red stars) and $28$ (black
  triangles).  The analytical results are given by the solution
  (\ref{steadySol}) with no fitting parameter and shown as solid lines
  for all values of $\kappa$. In all simulations, we use
  $\delta=1/N$.}
\label{fig:Numerical-and-analytical}
\end{figure}

The stationary state solution for Eq.~(\ref{diffusionEq}) can be readily
obtained by imposing that $\partial \rho/\partial t=0$, and both $\rho(x)$ and
$\partial\rho/\partial x$ go to zero as $x\rightarrow\pm\infty$,
\begin{equation}
  \rho_{\text{st}}(x)=\frac{1}{1+e^{\beta[\phi(x)-\mu]}},
  \label{steadySol}
\end{equation}
where $\mu$ is an integration constant. This solution corresponds to
the Fermi-Dirac distribution, with $\mu$ as the chemical potential.
It matches exactly the distribution obtained by making the entropy
(\ref{entropy}) an extreme, where the parameter $\mu$ can be
determined by the normalization,
\begin{equation}
  \int_{-\infty}^{\infty}\frac{\rho(x,t)}{\delta}dx=N.
  \label{normalization}
\end{equation}
As shown in Fig.~\ref{fig:Numerical-and-analytical}, the solution
(\ref{steadySol}) is in excellent agreement with the spatial profiles
of time-averaged occupancy obtained from numerical simulations for
distinct forms of the potential, namely, $\phi(x)=\kappa {|x|}^{n}$,
$n=1,\,2,\,3$ and $4$, and different values of $\kappa$ (or
temperature). As the strength of the confining potential increases (or
the temperature decreases), the maximum occupancy density at the
center of the potential approaches unity, $\rho_{\text{st}}\approx 1$,
and the peak in the profile becomes narrower. At this point, since the
density can not increase further, any additional confinement leads to
more sites with a maximum average occupancy, resulting in the
characteristic step shape of the Fermi-Dirac distribution.

\begin{figure}[t]
\includegraphics[width=\columnwidth]{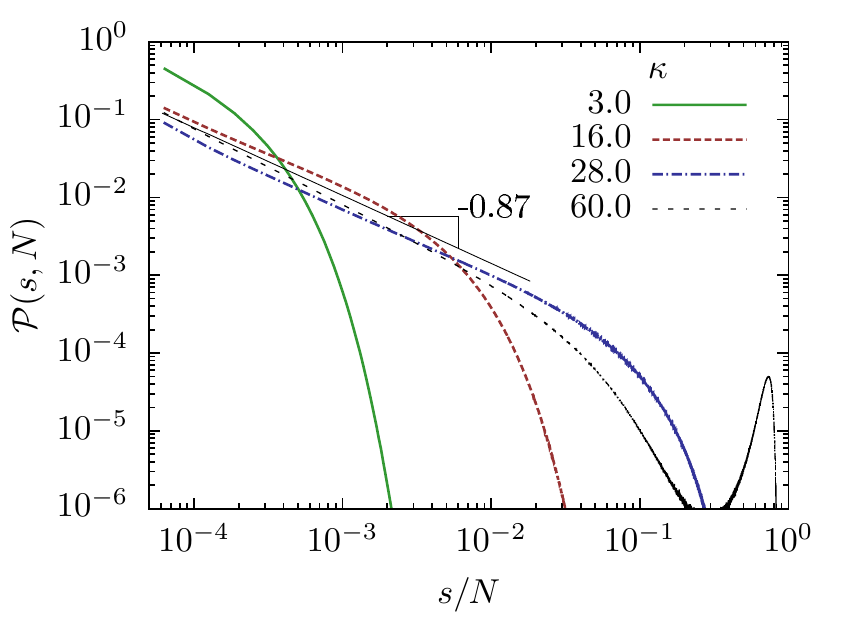}
\caption{Stationary-state distributions of avalanche sizes $s$ for the
  confined two-state sandpile model in 1D. The number of grains is
  $N=16000$ and $\beta=1$ for the numerical simulations. For
  $\kappa\approx 28$ the distribution displays power-law behavior,
  $P(s)\sim s^{-\alpha}$, followed by a cutoff of the form
  $\exp(-s^{2})$ at the order of the system size. The least-squares
  fit to the data of a power-law in the scaling region gives
  $\alpha=0.87\pm 0.02$.}
\label{fig.avalance_kdep}
\end{figure}

The confining potential substantially changes the way grains hop to
the nearest empty site. The average distribution of avalanches with
size $s$ is shown in Fig.~\ref{fig.avalance_kdep} for the case of
parabolic confinement and different values of $\kappa$. Here, a hop
from site $i$ to $j$ corresponds to an avalanche of size $|j-i|$. As
depicted, the distribution is an exponential decay for small values of
$\kappa$, in agreement with the derivation for the two-state sandpile
model without confinement \cite{Carlson1990}. By increasing $\kappa$
large avalanche become more probable, since the confinement favors the
occurrence of large clusters of grains near the center of the
potential. For a critical value of $\kappa\approx28$ the average
occupancy near the center of the potential approaches 1, and the
avalanche size distribution exhibits a power-law characteristics for a
wide range of sizes. More precisely, as indicated in
Fig.~\ref{fig.avalance_kdep}, $\mathcal{P}(s)\sim s^{-\alpha}$, with
$\alpha=0.87 \pm 0.02$. Further increase in the confinement parameter
$\kappa$ eventually leads to the occurrence of a very large cluster,
with near all the grains, located at the center of the symmetrical
potential, and, as a result, a pronounced peak for $s/N\sim 1$ become
evident in the avalanche size distribution. This corresponds to
avalanches spanning from one side of the system to the other, passing
through the center of the symmetrical potential.

\begin{figure}[t]
\includegraphics[width=\columnwidth]{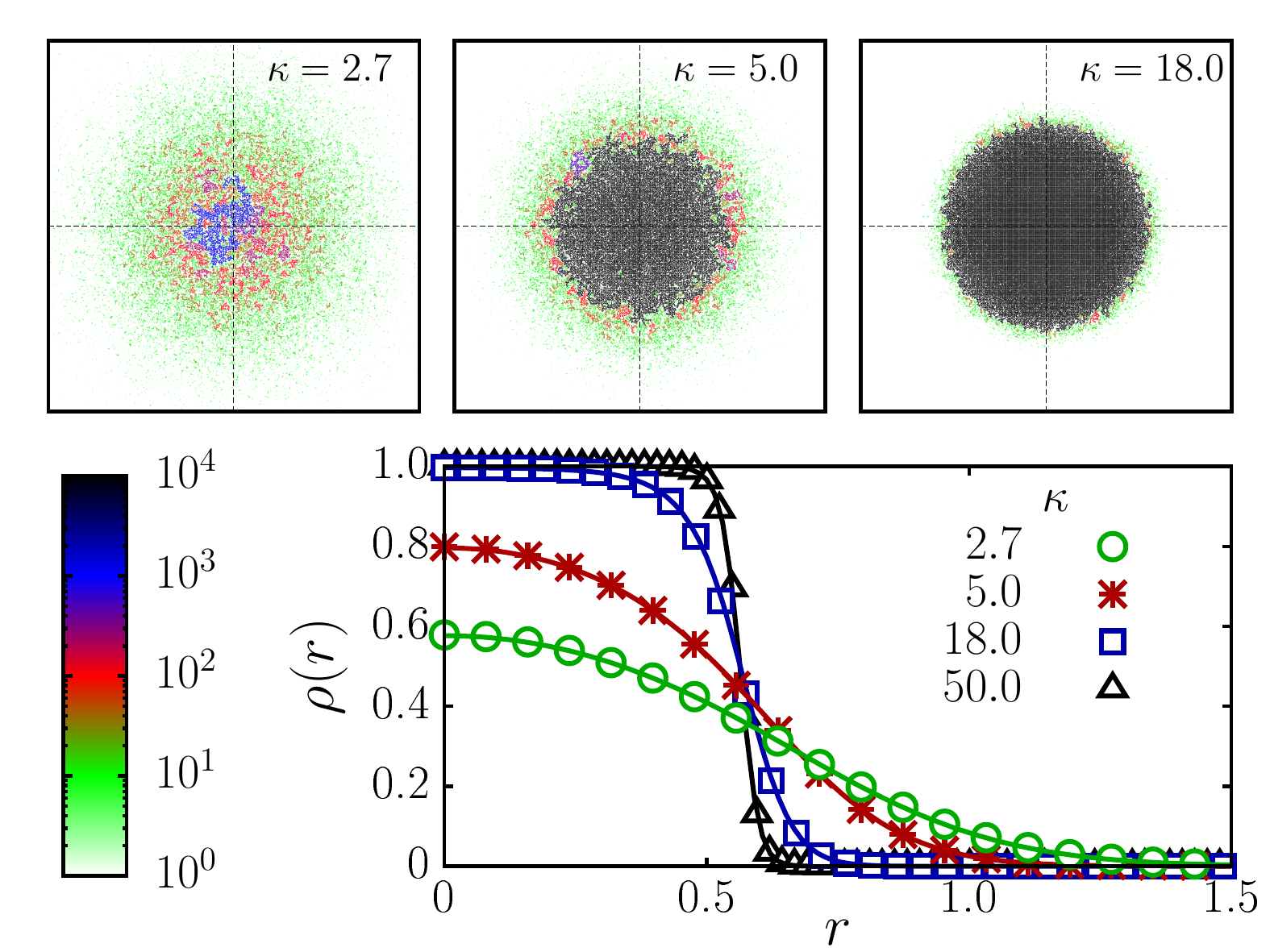}
\caption{On the top are snapshots of the grain positions for different
  values of $\kappa$ at the stationary state. Different colors
  correspond to distinct clusters and the color code in the bar
  indicates the cluster size, which increases from white to black.  On
  the bottom are the stationary-state profiles of the average
  occupancy, $\rho(r)$, for two-state two-dimensional sandpiles
  confined by a parabolic potential. The system size is $N=100000$ and
  different curves correspond to distinct values of the potential
  strength $\kappa$. The solid lines correspond to the analytical
  solutions (\ref{steadySol_2d}) without fitting parameters,
  calculated for different values of $\kappa$ and $\beta=1$. In all
  simulations, we use $\delta=1/\sqrt{N}$.}
\label{fig:Stationary-state-average2D}
\end{figure}

Next we extend our results to two-dimensional systems. In this case, a
grain at position $\mathbf{r}_i=(x_{i},y_{i})$ moves in a randomly
selected direction until it finds the nearest empty site.  The
transition is then accepted or not following the same Metropolis
algorithm previously described for the 1D case, but now with a
confining parabolic potential of the form,
$\phi(\mathbf{r}_i)=\kappa(x_{i}^{2}+y_{i}^{2})$.
Figure~\ref{fig:Stationary-state-average2D} shows the radial profile
of the time average occupancy in 2D, $\rho(r)$, obtained from
numerical simulations for different values of the strength of the
confining potential $\kappa$. The qualitative behavior of the system
is the same as in 1D, namely, the stronger the confining potential,
the narrower the profile with the maximum occupancy at the center of
the potential approaching unit. Further increasing $\kappa$, the
occupancy saturates at $\rho\sim1$ and the profile becomes broader,
resembling a step function. Also shown in
Fig.~\ref{fig:Stationary-state-average2D} are typical snapshots of the
grain positions for the same values of $\kappa$, colored according to
the size of the clusters they belong to. If the confinement is weak,
all sizes of clusters are present, with larger clusters located at the
center of the potential. As $\kappa$ increases, larger and more
compact clusters are favored at the center of the potential, tending
to a limit where most of the grains belong to a single, compact
cluster with an irregular surface.
 
As for the one-dimensional case, the results in
Fig.~\ref{fig:Stationary-state-average2D} computed for distinct
confinement strengths show that the average radial profiles of
occupancy in 2D are perfectly consistent with the Fermi-Dirac
distribution,
\begin{equation}
  \rho_{\text{st}}(r)=\frac{1}{1+e^{\beta[\phi(r)-\mu]}},
  \label{steadySol_2d}
\end{equation}
but now subjected to the normalization condition,
\begin{equation}
  \int_{0}^{2\pi}\int_{0}^{\infty}\frac{\rho(r,t)}{\delta^2}rdrd\phi = N \quad \Rightarrow \quad 
  \mu=\frac{1}{\beta}\ln\left(e^{\beta\frac{N\delta^2\kappa}{\pi}}-1\right).
  \label{normalization_2d}
\end{equation}
This excellent agreement between simulations and the Fermi-Dirac
distribution suggests that in two-dimensions the system satisfies the
generalization of the FPE~(\ref{FPE_Eq}) to higher dimensions, which
is of the form,
\begin{equation}
  \frac{\partial\rho}{\partial t}=
  \nabla \cdot \left[\Omega(\rho)\nabla\rho\right]
  -\nabla \cdot \left[\mathbf{A}(\mathbf{r})\Psi_(\rho)\right],
  \label{FPE_Eq_2d}
\end{equation}
where $\mathbf{A}(r)=-\nabla \phi(\mathbf{r})$, with the condition
(\ref{d2gdp2}) still valid in 2D.

\begin{figure}[t]
\includegraphics[width=\columnwidth]{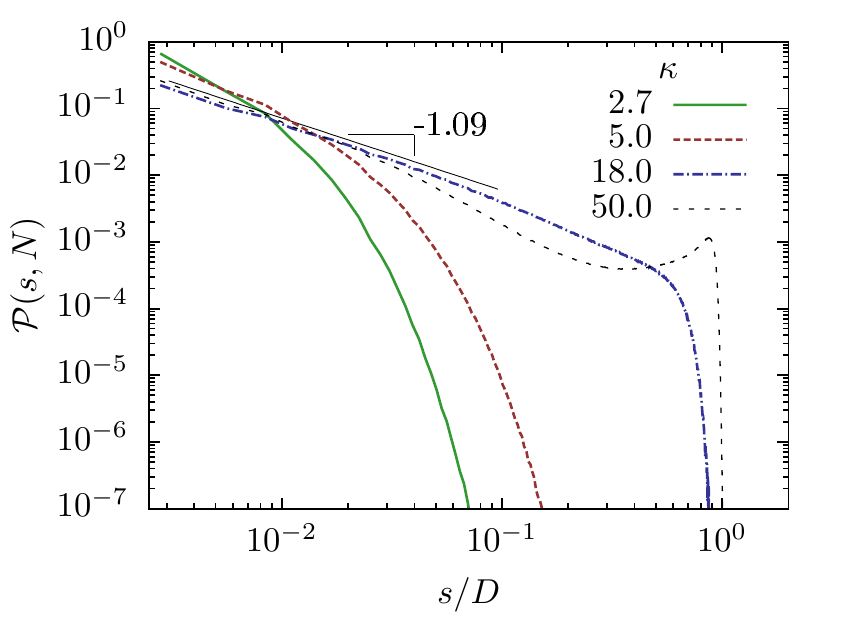}
\caption{Stationary-state distributions of avalanche sizes $s$ for the
  confined two-state sandpile model in 2D. The number of grains is
  $N=100000$ and $\beta=1$ for the numerical simulations. For
  $\kappa\approx 18$ the distribution displays power-law behavior,
  $P(s)\sim s^{-\alpha}$, followed by a cutoff of the form
  $\exp(-s^{2})$ at the order of the characteristic system size,
  $D\approx 2\sqrt{N/\pi}$. The least-squares fit to the data of a
  power-law in the scaling region gives $\alpha=1.09\pm 0.04$.}
\label{fig.avalance_kdep2D}
\end{figure}

Figure \ref{fig.avalance_kdep2D} depicts the avalanche size
distribution for the two-dimensional system. Here $s$ corresponds to
the number of sites visited in the fixed chosen direction until an
empty site is found and the transition is accepted. As shown in Fig.
\ref{fig.avalance_kdep2D}, the avalanche size distribution changes
from an exponential decay for $\kappa=2.7$ to a distribution with a
well defined peak near the size of the system for $\kappa=50.0$. In
2D a characteristic size for the avalanches is the diameter of the
system $D\approx 2 \sqrt{N/\pi}$. Our numerical results show that,
for a critical value of the confining parameter, $\kappa\approx18$,
the presence of the confining potential leads to an avalanche size
distribution that obeys a power law for small avalanche sizes,
$P(s)\sim s^{-\alpha}$, with an exponent $\alpha=1.09\pm 0.04$.

In summary, here we studied the effect of a confining potential on the
behavior of a two-state sandpile model in one and two dimensions. A
continuum nonlinear diffusion equation could be derived from the
microdynamical description of the model that is shown to be perfectly
consistent with the transport of grains observed from numerical
simulations. This equation, besides displaying singularities in both
the diffusion and drift terms, has a stationary-state solution for the
spatial profiles of average occupancy of grains that maximizes the
Fermi-Dirac entropy. Moreover, our results show that the introduction
of a confining potential to the two-state sandpile model, if properly
tuned, can lead to power-law behavior in the distribution of
grain-jump sizes. These results are rather surprising since 1D
systems usually do not display non-trivial critical states nor
power-law behavior. They can be explained in terms of the
non-homogeneity introduced by the confining potential and the complex
fluctuations due to the singular-diffusion dynamics. The extension to
two-dimensions reveals that the strongly nonlinear features of the
system together with the intrinsic exclusion mechanism present in the
model also lead to the Fermi-Dirac distribution for the occupancy
profiles. Power-law distributions of avalanches sizes are also
observed in 2D at critical values of the intensity of the confining
potential.

We thank the Brazilian agencies CNPq, CAPES, and FUNCAP for financial support.

\end{document}